\documentstyle[12pt,aasms4]{article}

\received{XX XXXXX 1996}
\accepted{YY YYYYY 1996}
\journalid{ZZ}{ZZ ZZZZZZZ 199Z}
\articleid{uu}{uu}

\lefthead{Mori et al.}
\righthead{The Evolution of Dwarf Galaxies}

\begin{document}

\title{THE EVOLUTION OF DWARF GALAXIES WITH STAR FORMATION IN
OUTWARD PROPAGATING SUPER SHELL}

\author{Masao Mori\altaffilmark{1,}\altaffilmark{2},
Yuzuru Yoshii\altaffilmark{3,}\altaffilmark{4},
Takuji Tsujimoto\altaffilmark{5}
and Ken'ichi Nomoto\altaffilmark{1,}\altaffilmark{4}}

\altaffiltext{1}{Department of Physics, School of science,
Nagoya University, Chikusa-ku, Nagoya 464-01, Japan}
\altaffiltext{2}{The Institute of Physical and Chemical Research(RIKEN),
Wako, Saitama, 351-01,Japan}
\altaffiltext{3}{Institute of Astronomy, Faculty of science, 
University of Tokyo, Mitaka, Tokyo 181, Japan}
\altaffiltext{4}{Research Center for the Early Universe, School of science,
University of Tokyo, Bunkyo-ku, Tokyo 113, Japan}
\altaffiltext{5}{National Astronomical Observatory, Mitaka, Tokyo 181,
Japan}
\altaffiltext{6}{Department of Astronomy, School of science, 
University of Tokyo, Bunkyo-ku, Tokyo 113, Japan}

\authoremail{mori@astron.s.u-tokyo.ac.jp}

\begin{abstract}

We simulate the dynamical and chemical evolution of a dwarf galaxy
embedded in a dark matter halo, using a three-dimensional $N$-body/SPH 
simulation code combined with stellar population synthesis.  
The initial condition is adopted in accord with a $10^{10}M_\odot$ virialized
sphere in a $1\sigma$ CDM perturbation which contains 10\% baryonic mass.  
A supersonic spherical outflow is driven by the first star burst near 
the center of the galaxy and produces an expanding super shell in which 
stars are subsequently formed.  Consecutive formation of stars in the 
expanding shell makes the stellar system settled with the exponential 
brightness profile, the positive metallicity gradient, and the inverse 
color gradient in agreement with observed features of dwarf galaxies.  
We therefore propose that the energy feedback via stellar winds and 
supernovae is a decisive mechanism for formation of less compact, small
systems like dwarf galaxies.
\end{abstract}


\section{Introduction}
Our understanding of how galaxies originate and distribute on large scales
in the universe has greatly improved in the last two decades.  
While the standard model of hierarchical galaxy clustering 
\markcite{WR1987}(White \& Rees 1978) has been successful in explaining 
the clustering pattern of galaxies revealed by redshift surveys, 
it predicts a large number of low-mass galaxies ($L<10^{10}L_{\odot}$)
beyond that estimated from the observed luminosity function of galaxies
(\markcite{WF1991}White \& Frenk 1991;
 \markcite{CAFNZ1994} Cole {\it et al.} 1994).  
The hierarchical model should therefore involve some mechanism which
suppresses the formation of such small galaxies.  Main mechanisms so 
far proposed include an energy feedback from supernovae that prevents 
the collapse of a forming galaxy (Dekel \& Silk 1986; Lacey \& Silk 1992) 
and a photoionization by ultraviolet background radiation that keeps 
the gas hot and unable to collapse (Dekel \& Rees 1987; Efstathiou 1992).

Significant body of new observations of nearby dwarf galaxies has reveald 
a web of filaments, loops and expanding super giant shells which are 
imprinted in the ionized gas around individual galaxies
(\markcite{MFD1992}Meurer, Freeman \& Dopita 1992;
 \markcite{MHW1995}Marlowe, Heckman \& Wyse 1995;
 \markcite{H1996}Hunter 1996).  Since the traces of energetic winds 
are oriented from supernovae or massive stars, it is evident that the
heat input from them greatly affects the dynamics of small galaxies.
This feedback of energy into the interstellar medium must play a 
decisive role in the early stage of galaxy evolution when star formation 
rate is expected to be much higher.  

Dekel \& Silk (1986) showed that the supernova feedback mechanism nicely 
accounts for the observed correlations between metallicity, color and 
luminosity of galaxies (see also Vader 1986; Yoshii \& Arimoto 1987).  
There is however a clear distinction in structural and chemical quantities 
bewteen dwarf ellipticals (dEs) and normal ellipticals in spite of 
their morphological similarity.  The central concentration of dEs is 
relatively low and their luminosity profiles are best fitted by an 
exponential function, whereas the profiles of normal ellipticals are 
known to follow de Vaucouleurs' law 
(\markcite{FL1983}Faber \& Lin 1983;
 \markcite{BST1984}Binggeli, Sandage \& Tarenghi 1984;
 \markcite{IWO1986}Ichikawa, Wakamatsu \& Okamura 1986;
 \markcite{CB1987}Caldwell \& Bothun 1987).
Moreover, the color of many dEs becomes redder towards outer radii of 
the system (Vader {\it et al.} 1988; Kormendy \& Djorgovski 1989; 
Chaboyer 1994), and this trend of color gradient is clearly opposite 
to normal galaxies.  
The origin of these striking features of dEs remains yet to be explained 
(for a review see Ferguson \& Binggeli 1994). In particular, no attempts 
have ever been made to examine whether the supernova feedback mechanism 
is viable also in this context.

In this paper, we use three dimensional simulation code with a 
cosmologically motivated initial condition and investigate the formation 
and evolution of a dE galaxy taking into account the dynamical responses
of the system from supernova-driven winds.  Our simulation shows that 
such winds propagating outwards from inside the system collide with the
infalling gas and produce the super shell in which stars are formed.  
This specific process of star formation turns out to reproduce the 
observed features of dEs, and therefore the heating by supernovae proves 
to be an ideal suppressing mechanism against the efficient formation of 
low-mass galaxies in the hierachical clustering model.

\section{Numerical Method}

Our simulation uses a hybrid $N$-body/hydrodynamics code which is 
applicable to a complex system consisting of dark matter, stars and
gas.  The gas is allowed to form stars and is subject to 
physical processes such as the radiative cooling and the energy feedback 
from supernovae and massive stars.  The cooling rate of the gas is
calculated assuming the primordial composition, and the effect of
photoionization by ultraviolet background radiation is ignored for 
simplicity.  Chemical and photometric evolution of the system can 
also be simulated by this code.  The collisionless dynamics for dark 
matter particles and stars is treated by the $N$-body method and the 
gas dynamics by the method of smoothed particle hydrodynamics (SPH) 
(\markcite{HK1989}Hernquist \& Katz 1989; \markcite{M1992}Monaghan 1992).
Our numerical technique is essentially similar to that adopted by
\markcite{S1996}Steinmetz (1996).  We only briefly describe 
how to calculate the self-gravity, star formation, and energy feedback.
The details will be given in a forthcoming paper 
(\markcite{MYTN1996c}Mori {\it et al.} 1996b). 

Self-gravity calculations are run on the hardware GRAPE-3AF 
(\markcite{GRAPE1990}Sugimoto {\it et al.} 1990) by using the ``Remote-GRAPE'' 
system.  This remote system is newly developed in order to allow an access
to the GRAPE-3AF from local workstations which are not physically connected 
to the host workstation.  Thus, self-garvity calculations can be performed
in parallel with other calculations, so that the calculation time is 
considerably shortened.  The performance analysis of this system is 
reported by 
\markcite{NMN1996}Nakasato {\it et al.} (1996) and 
\markcite{MYTN1996c}Mori {\it et al.} (1996b).

Stars are assumed to form in rapidly cooling, Jeans unstable and converging 
regions at a rate which is inversely proportional to the local dynamical 
time (\markcite{K1992}Katz 1992;
\markcite{NW1993}Navarro \& White 1993;
\markcite{SM1994}Steimetz \& M\"uller 1994).
When a star particle is formed, we identify this with approximately 
$10^4 $ single stars and distribute the associated mass of the star 
particle over the single stars according to Salpeter's (1955) initial 
mass functiion.  The lower and upper mass limits are taken as
$m_l=0.1M_\odot$ and $m_u=50M_\odot$, respectively. 

Our SPH algorithm for treating the energy feedback from massive stars
is a more physically motivated one and is different from those adopted
by previous authors
(\markcite{K1992}Katz 1992; \markcite{NW1993}Navarro \& White 1993;
 \markcite{MH1994}Mihos \& Hernquist 1994). 
When a star particle is formed and identified with a stellar assemblage 
as described above, stars more massive than 8 $M_{\odot}$ start to explode 
as Type II supernovae (SNe II) with the explosion energy of $10^{51}$ ergs
and their outer layers are blown out with synthesized metals leaving the 
remnant of 1.4 $M_{\odot}$.  We can regard this assemblage as continuously
releasing the energy at an average rate of 
$8.44\ 10^{35}$ ergs sec$^{-1}$ per star during the explosion period 
from $t(m_u)=5.4\ 10^6$ yrs until $t(8M_\odot)=4.3\ 10^7$ yrs 
where $t(m)$ is the lifetime of a star of mass $m$.   
Prior to the onset of SN explosions, however, their progenitors 
develope stellar winds and also release the energy of $10^{50}$ ergs 
into the interstellar medium at an average rate of 
$7.75\ 10^{34}$ ergs sec$^{-1}$ per star.  Consequently, once a new 
star particle is formed, the energy from stellar winds is supplied to the
gas particles within a sphere of radius $R_{snr}$, and the energy, 
metals and material from SNe II are subsequently supplied to the same
region.  The radius $R_{snr}$ is set equal to the maximum extension of the 
shock front in the adiabatic phase of supernova remnant and is given by 
$R_{snr}=32.9 E_{51}^{\;1/4}\,n^{-1/2}$ pc 
(\markcite{SS1979}Shull \& Silk 1979) where $E_{51}$ is the released energy 
in units of $10^{51}$ergs and $n$ is the number density of the gas in units 
of cm$^{-3}$ which surrounds the star particle.  The gas within $R_{snr}$ 
remains adiabatic until multiple SN phase ends at $t(8M_\odot)$, and then 
it cools according to the adopted cooling rate of the gas.
We compute the chemical evolution using the new calculations 
of stellar nucleosynthesis products (Tsujimoto {\it et al.} 1995).

\section{Simulation Result}

Following a standard model of the cold dark matter (CDM) universe
($\Omega_0=1$, $H_0=50$ km$\,$sec$^{-1}$Mpc$^{-1}$), we consider
a less massive protogalaxy as a gas sphere with mass of $10^{9}\,M_{\odot}$
embedded in a $1 \sigma$ density peak having a total mass of 
$10^{10}M_{\odot}$ with a baryon to dark matter ratio equal to 1/9. 
The distribution of dark matter halo is assumed to have a King profile 
with the central concentration index of $c=1$.   This two-component system 
is made to settle in a virial equilibrium from which the gas temperature 
and the velocity dispersion of dark matter are estimated as an initial 
condition.   Our simulation uses $10^4$ particles for each of gas and dark 
matter particles.  The gravitational softening parameter is adopted as 
78.9 pc for gas particles and 36.6 pc for collisionless particles.

As soon as we start a simulation, the gas in the central region of
the protogalaxy rapidly cools and begins to contract owing to the 
self-gravity of dark matter and gas.  When the gas temperature becomes 
close to $10^4 K$ and stops decreasing, a quasi-isothermal contraction 
is established.   A further increase of the gas density causes a burst
of star formation in the central region.  Thereafter, as massive stars
explode as SNe II, the surrounding gas aquires the thermal energy and
the gas temperature rises up to about $10^6$K.  At the same time, the
gas is gradually polluted with synthesized metals from SNe II.  About
5\% of the initial gas mass is used up in this formation of the 
first generation stars.  

The shock waves propagate outwards and the supernova-driven spherical 
outflow occurs from inside.  This outflow collides with the infalling
gas and the high-density super shell is eventually formed.  While the gas 
is continuously swept up by the super shell, the gas density further
increases due to the enhanced cooling rate in the already dense shell.
Then the intense formation of stars begins within the super shell,
and subsequent SN explosions further accelerate the outward expansion 
of the shell.  Star formation continues in the expanding shell for
about $10^8$ yrs until the gas density in the shell becomes too low to
form new stars.  About 26\% of the initial gas mass is turned into stars
in this stage.  The remaining gas in the shell is blown out to the 
intergalactic space at supersonic speed.  
The ejected gas has already enriched to the yield value 
$y_Z\approx Z_\odot$ having the metal abundance of 
$\log Z/Z_\odot \sim 0.0$.
Figure 1 shows the ring-like 
distribution of gas particles and newly born star particles near the 
$X-Y$ sectional plane at the elapsed time of $\sim 2\ 10^7$ yrs in the 
simulation.   It is evident from this figure that the star-forming site
is well confined in the shell.

In such a way, a total of about 31\% of the initial gas mass has
turned into stars before the dwarf galaxy is formed.  The baby stars 
initially have the velocity vectors of the gas from which the stars are 
formed.  Therefore, the first generation stars have zero systematic 
velocity, but the later generation stars has a large outward 
radial velocity component. The oscillation of swelling and contraction of 
the system continues for several $10^8$ yrs, and the system becomes
settled in a quasi-steady state in $3\ 10^9$ yrs.  
The resulting stellar system forms a loosely bound virialized system
due to the significant mass loss and has a large velocity dispersion 
and a large core.  Consequently the surface mass distribution is 
approximately exponential (Figure 2a) and differs from the de Vaucouleur's 
profile which is more concentrated towards the galaxy center.
In order to enable a more direct comparison 
with the observation, we have computed the photometric evolution up to
10 Gyrs based on the method of stellar population synthesis, using the 
updates of stellar evolutionary tracks compiled by Kodama \& Arimoto 
(1996).  The resulting surface $B$-band brightness distribution at 10 Gyrs 
is obviously exponential (Figure 2b).  The effective radius within
which a half of the total light is contained is 1.42 kpc.
The integrated blue luminosity of the system is $M_B=-14.5$ mag.

Stars are formed for the most part before the gas is fully polluted to 
the yield value $y_Z\approx Z_\odot$ of the synthesized metals.  
The average metal abundance of the stars in the system is as low as 
$\log Z/Z_\odot\sim -1.74$.  This metallicity is consistent with
a range covered by the observations, but is much lower than those
of normal galaxies (Dekel \& Silk 1986; Yoshii \& Arimoto 1987).  
One outstanding feature discovered by our simulation is that the radial 
distribution of metal abundance in this system has a {\it positive} 
gradient (Figure 2c) which is in sharp contrast to the observed negative 
gardient for massive galaxies (Carollo, Danziger \& Buson 1993).  
We note that the star-forming site moves outwards with the expanding shell 
and the gas in this shell is gradually enriched with synthesized metals 
from SNe II.  Stars of later generations are necessarily born at larger radii 
with larger metallicities, leading to emergence of the positive metallicity 
gradient in the resulting stellar system.

Since the $V-K$ color sensitively traces the metallicity of underlying
stellar population (Yoshii \& Arimoto 1991), we calculate the radial 
distribution of the integrated $V-K$ color (Figure 2d), and the result 
is consistent with the observed trend of the inverse color gradient for 
dwarf galaxies (Vader {\it et al.} 1988; Kormendy \& Djorgovski 1989; 
Chaboyer 1994).

\section{Summary \& Discussion}

A three-dimensional $N$-body/SPH simulation code, combined with stellar 
population synthesis, is used to follow the dynamical and chemical 
evolution of a dwarf protogalaxy with $10^{10}M_\odot$ (baryonic/dark=1/9) 
which originates from a $1\sigma$ CDM perturbation.  This less massive 
galaxy receives significant dynamical responses from the heat input by 
stellar winds and supernovae.  

The first star burst near the center of the system produces a supersonic 
spherical outflow of the gas.  This outflow collides with the infalling 
gas and gives rise to an expanding dense shell.  Then, stars begin to 
form in the expanding shell with its site propegating outwards with 
the shell.  We find from the simulation that this consecutive process
of star formation creates the exponential brightness profile and the 
inverse color gradient of the system in agreement with the observations 
of dwarf galaxies. 

\markcite{A1994}Athanassoula (1994) performed one-dimensional simulations 
of the dynamical evolution of dE galaxies including the energy feedback 
from supernovae.  The models without dark halo are shown to give a better
agreement with observations than those with dark halo.  Our more realistic,
three-dimensional simulations however indicate that the dark halo is 
necessary and plays a vital role to form the bound stellar system, 
otherwise the system is blown out to disrupt completely.  

In general, the color gradient of galaxies is created by the gradient
in either metallicity or age of the underlying stellar population.  
Simple models of chemical evolution of galaxies usually predict the 
negative metallicity gradient which corresponds to the color becoming 
redder towards the galaxy center.  Since dwarf galaxies have the inverse
color gradient, Vader {\it et al.} (1988) were led to interpret this
observed trend in terms of the positive age gradient.  We note however
that stars with very low metallicities must have been formed on very 
short timescales and therefore no appreciable age difference results.

The above puzzling situation indicates that previous results based on 
simple models of chemical evolution can not be applied to small systems 
like dwarf galaxies.   We demonstrate in this paper that dynamical 
modelling is the only proper way to investigate the evolution of dawrf 
galaxies.  Successful reproduction of their basic features in our 
simulation suggests that the stellar energy feedback mechanism is indeed 
a likely mechanism against the efficient formation of low-mass galaxies 
in the CDM universe.

\acknowledgments
We are grateful to T. Shigeyama and M. Chiba for many fruitful discussions,
to T. Kodama for providing us the tables of population synthesis prior to 
the publication, and to N. Nakasato for preparing the Remote-GRAPE library.  
This research has been supported in part by the Grant-in-Aid for Scientific 
Research (05242102, 06233101) and Center-of-Excellence Research (07CE2002) 
of the Ministry of Education, Science, and Culture in Japan.

\clearpage

\figcaption[figure1.eps]{Distribution of gas and star particles near 
the $X-Y$ sectional plane ($|X|\leq 0.5$ kpc,
$|Y|\leq 0.5$ kpc, $|Z|\leq 0.05$ kpc).  {\it Left panel:} snapshot
for gas particles at the elapsed time of $2.17\ 10^7$ yrs.  
{\it Right panel:} snapshot for newly born star particles between 
$2.05\ 10^7$ yrs and $2.17\ 10^7$ yrs. 
\label{fig1}}  

\figcaption[figure2.eps]{Various simulated quantities as a function 
of radial distance away from the center of the bound stellar system 
at the elapsed time of 10 Gyrs.  All the radial profiles are plotted 
against the radius in units of kpc (solid lines) or a quartic root 
of the radius (dotted lines).  
(a) Surface stellar mass density.  
(b) Surface $B$-band brightness.
(c) Logarithmic stellar metal abundance. 
(d) Integrated $V-K$ color.
\label{fig2}}

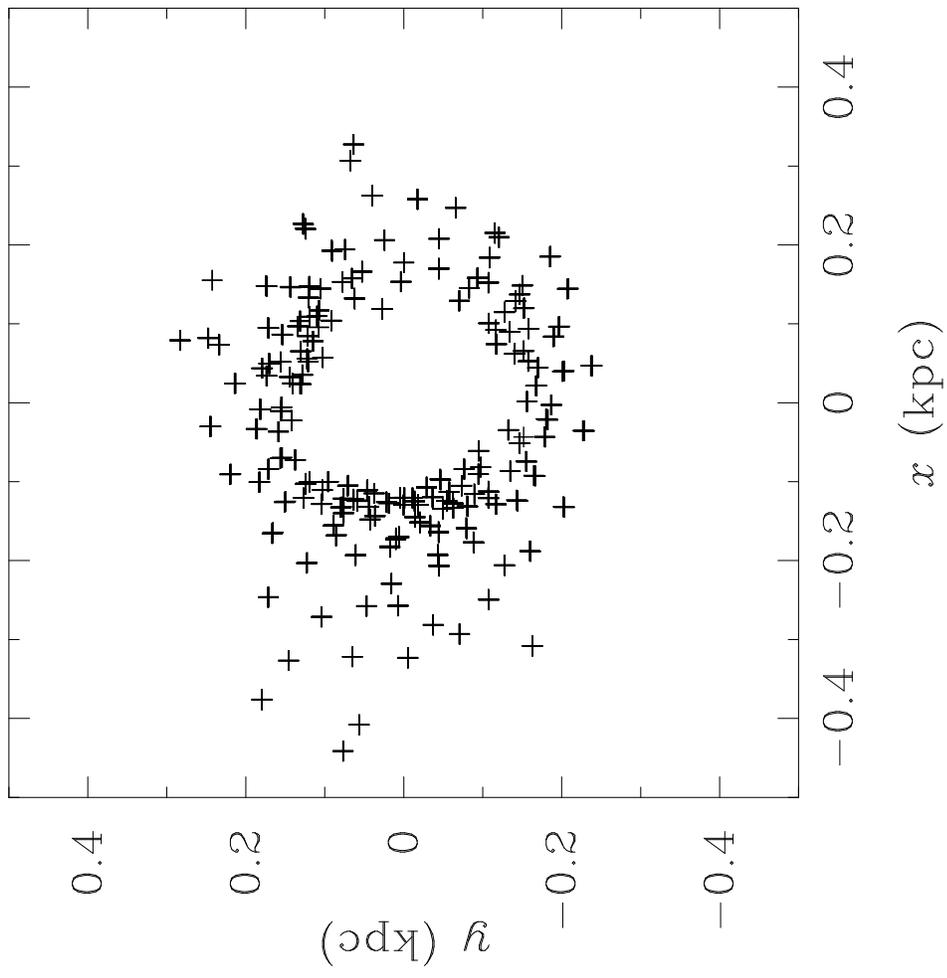
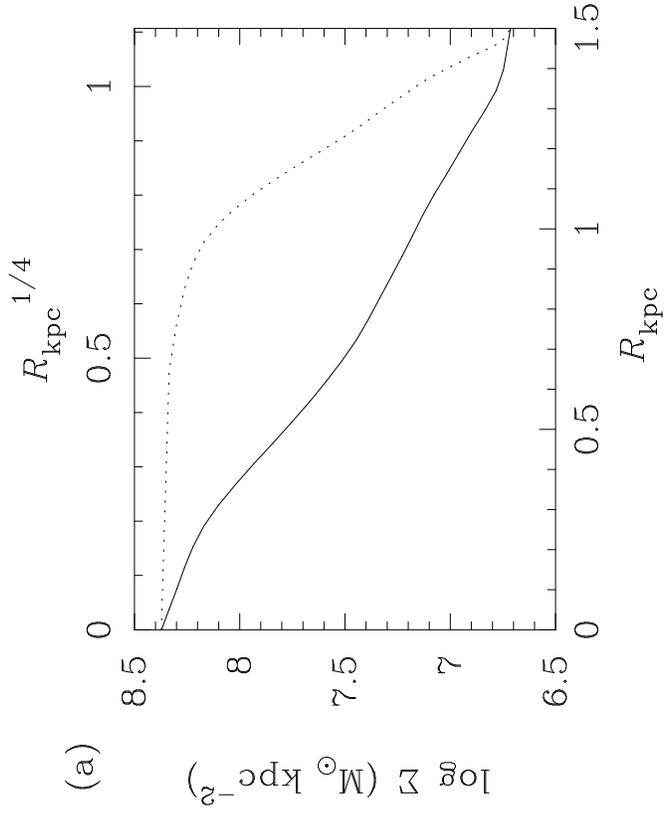
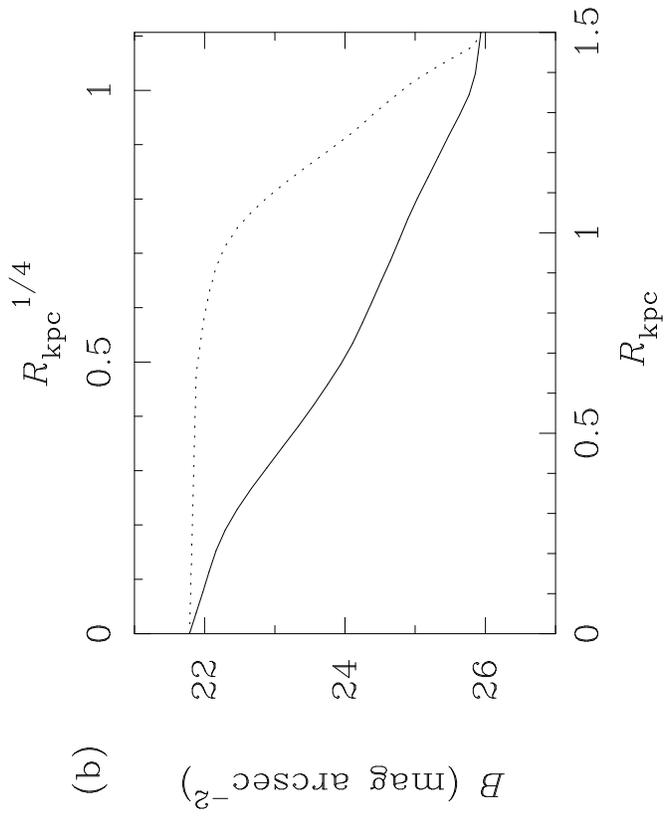
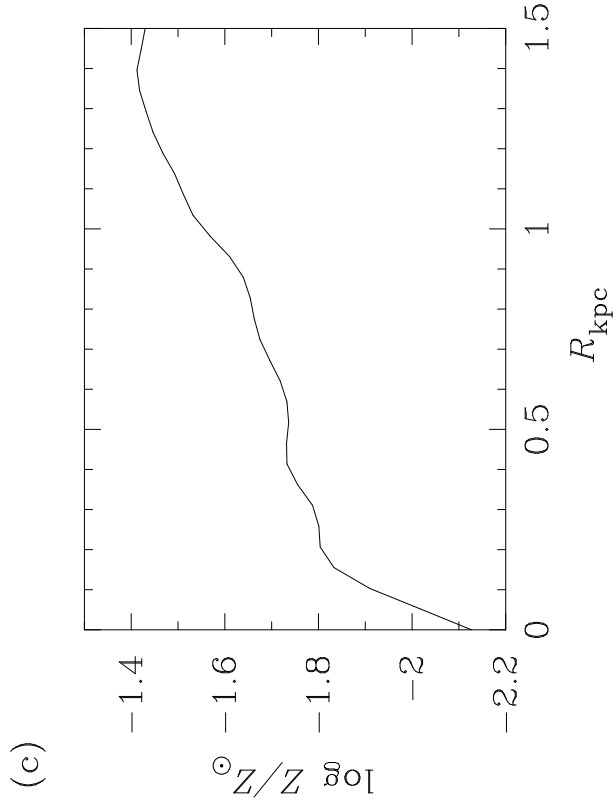
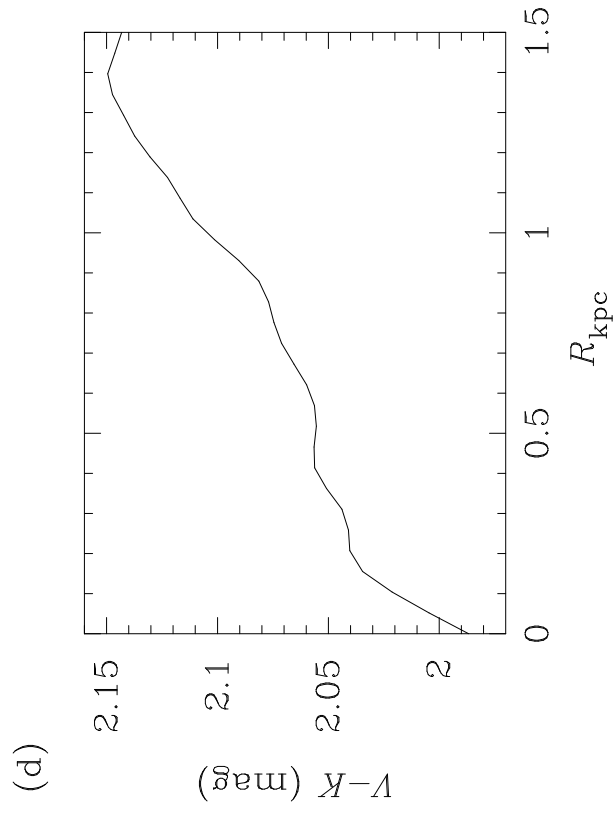
\end{document}